\documentclass[12pt,final]{IEEEtran}
\usepackage{amsmath, graphicx, amssymb, amsfonts, amsthm, subfigure, arydshln, setspace}

\newcommand{\graph}{{\cal   G}}

\newcommand{\bw}{{\bf{w}}}
\newcommand{\bu}{{\bf{u}}}
\newcommand{\bp}{{\bf{p}}}
\newcommand{\less}{\le}
\newcommand{\greater}{\ge}
\newcommand{\identity}{I}
\newcommand{\zero}{O}
\newcommand{\cR} {{\mathcal{R}}}
\newcommand{\cC}{{\cal{C}}}

\newcommand{\sinks}{{\mathcal T}}

\newcommand{\code}{\cC}

\newcommand{\sink}{t}

\newcommand{\sources}{{\mathcal S}}

\newcommand {\Graph} {{\mathcal G}}
\newcommand {\Edges} {{\cal E}}

\newcommand {\Nodes} {{\mathcal V}}
\newcommand {\F} {\mathbb F}
\newcommand {\DisPath} {\mathcal P}

\newcommand {\SourceNum}{s}
\newcommand {\NetSize} {|\mathcal {E}|}

\newcommand {\nchoosek}[2]{\left(\begin{array}{c}#1\\#2\end{array}\right)}

\newtheorem{theorem}{Theorem}

\newtheorem{lemma}{Lemma}
\newtheorem{proposition}{Proposition}

\begin{document}

\onecolumn
\doublespace

\title{Multiple-access Network Information-flow and Correction Codes$^*$}
\author{\thanks{$^*$ In other words, MANIAC codes.}
\thanks{$^{**}$ The first five authors had equal contribution to this work and they are named in alphabetical order.}Theodoros K. Dikaliotis$^{**}$, Tracey Ho$^{**}$,~\IEEEmembership{Member,~IEEE}, Sidharth Jaggi$^{**}$,~\IEEEmembership{Member,~IEEE}, Svitlana Vyetrenko$^{**}$, Hongyi Yao$^{**}$, Michelle Effros,~\IEEEmembership{Fellow,~IEEE}, J{\"o}rg Kliewer,~\IEEEmembership{Senior Member,~IEEE}, Elona Erez~\IEEEmembership{Member,~IEEE}
\thanks{T.K. Dikaliotis, T. Ho, M. Effros and S. Vyetrenko are with the Department of Electrical Engineering, California Institute of Technology,
          Pasadena, CA 91125, USA, e-mail: \{tdikal,tho,effros\}@caltech.edu and svitlana@acm.caltech.edu respectively.}
\thanks{S. Jaggi is with the Department of Information Engineering, The Chinese University of Hong Kong, Shatin N.T., Hong Kong, e-mail:sidjaggi@gmail.com.}
\thanks{H. Yao was with Tsinghua University, Beijing, China. He is now with the California Institute of Technology, Pasadena, CA 91125, USA, e-mail: yaohongyi03@gmail.com.}
\thanks{J. Kliewer is with the Klipsch School of Electrical and Computer Engineering, New Mexico State University, Las Cruces, NM 88003-8001, USA, e-mail: jkliewer@nmsu.edu.}
\thanks{E. Erez was with the California Institute of Technology, Pasadena, California.  She is now with the School of Engineering \& Applied Science, Yale University, e-mail: elona.erez@gmail.com.}
\thanks{Subsets of the authors T.K. Dikaliotis, S. Vyetrenko, T. Ho, M. Effros and E. Erez were supported by subcontract \#069153 issued by BAE Systems National Security Solutions, Inc. and by the Defense Advanced Research Projects Agency (DARPA) and the Space and Naval Warfare System Center (SPAWARSYSCEN), San Diego under Contract No. N66001-08-C-2013, AFOSR under Grant 5710001972, Caltech's Lee Center for Advanced Networking, and NSF grant CNS-0905615. The work of S. Jaggi was supported by the RGC GRF grants 412608 and 412809, the CUHK MoE-Microsoft Key Laboratory of Human-centric Computing and Interface Technologies, the Institute of Theoretical Computer Science and Communications, and Project No. AoE/E-02/08 from the University Grants Committee of the Hong Kong Special Administrative Region, China. The work of H. Yao was supported by the National Natural Science Foundation of China Grant 61033001 and 61073174, the National Basic Research Program of China Grant 2007CB807900 and 2007CB807901, the Hi-Tech research \& Development Program of China Grant 2006AA10Z216. The work of J. Kliewer was supported by NSF grant CCF-0830666.}}
\maketitle

\begin{abstract}This work considers the multiple-access multicast error-correction scenario over a packetized network with  $z$ malicious edge adversaries. The network has min-cut $m$ and packets of length $\ell$, and each sink demands all information from the set of sources $\sources$.  The capacity region is characterized for both a ``side-channel" model (where sources and sinks share some random bits that are secret from the adversary) and an ``omniscient" adversarial model (where no limitations on the adversary's knowledge are assumed). In the ``side-channel" adversarial model, the use of a secret channel allows higher rates to be achieved compared to the ``omniscient" adversarial model, and a polynomial-complexity capacity-achieving code is provided. For the ``omniscient" adversarial model, two capacity-achieving constructions are given: the first is based on random subspace code design and has complexity exponential in $\ell m$, while the second uses a novel multiple-field-extension technique and has $O(\ell m^{|\sources|})$ complexity, which is polynomial in the network size.  Our code constructions are ``end-to-end'' in that all nodes except the sources and sinks are oblivious to the adversaries and may simply implement predesigned linear network codes (random or otherwise).  Also, the sources act independently without knowledge of the data from other sources.

\end{abstract}

\section{Introduction}
Information dissemination can be optimized with the use of network
coding. Network coding maximizes the network throughput in multicast
transmission scenarios~\cite{ACLY00}.  For this scenario,
it was shown in \cite{LYC03} that linear network coding suffices to
achieve the max-flow capacity from the source to each receiving node. An algebraic framework for linear network coding was presented in \cite{KM03}. Further, the linear combinations employed at network nodes can be randomly selected in a distributed manner; if the coding field size is sufficiently large the max-flow capacity is achieved with high probability \cite{Ho_etal06}.

\begin{figure}
 \centerline{\includegraphics[scale=0.3]{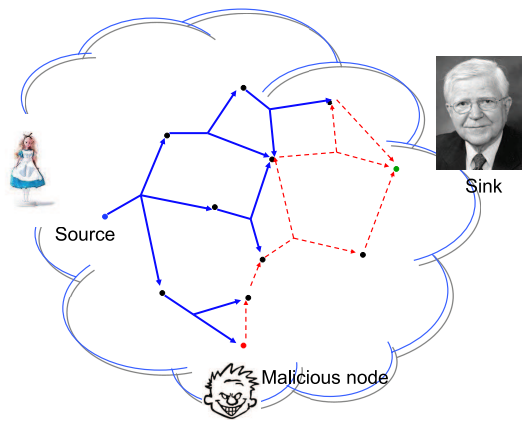}}
 \caption{Propagation of network errors via network coding. The
   action  of a single malicious node contaminates all incoming links
   of the sink node due to packet mixing at downstream nodes.}
 \label{fig:propagation}
\end{figure}

However, network coding is vulnerable to malicious
attacks from rogue users. Due to the mixing operations at internal
nodes, the presence of even a small number of adversarial nodes can
contaminate the majority of packets in a network, preventing sinks from decoding. In particular, an error on even a single link might propagate to multiple downstream links via network coding, which might lead to the extreme case in which  all incoming links at the sink are in error. This is shown in Fig.~\ref{fig:propagation}, where the action of a single malicious node contaminates all incoming links of the sink node due to packet mixing at downstream nodes.

In such a case, network error-correction (introduced in \cite{RaymondNECC2002}) rather than
classical forward error-correction (FEC) is required, since the former exploits the
fact that the errors at the sinks are correlated, whereas the latter assumes independent errors.

A number of papers e.g. \cite{YC06,CY06,SKK08} have characterized the set of achievable communication rates over networks containing hidden malicious jamming and eavesdropping adversaries, and given corresponding communication schemes. The latest code constructions (for instance~\cite{SKK08} and~\cite{Jaggi_etal08}) have excellent parameters -- they have low computational complexity, are distributed, and are asymptotically rate-optimal. However, in these papers the focus has been on single-source multicast problems, where a single source wishes to communicate all its information to all sinks.

In this work we examine the problem of multiple-access multicast, where multiple sources wish to communicate all their information to all sinks. We characterize the optimal rate-region for several variants of the multiple-access network error-correction problem and give matching code constructions, which have low computational complexity when the number of sources is small.

We are unaware of any straightforward application of existing single-source network error-correcting subspace codes that achieve the optimal rate regions. This is because single-source network error-correcting codes such as those of~\cite{Jaggi_etal08} and~\cite{SKK08} require the source to judiciously insert redundancy into the transmitted codeword; however, in the distributed source case the codewords are constrained by the independence of the sources.

\section{Background and related work}

For a single-source single-sink network with min-cut $C$, the capacity of the network
under arbitrary errors on up to $z$ links is given by
\begin{eqnarray}
\label{single-source-capacity}
R\le C-2z
\end{eqnarray}
and can be achieved by a classical end-to-end error-correction code over multiple disjoint paths from source to the sink. This result is a direct extension of the Singleton bound (see, {\it e.g.}, \cite{Roth06}). Since the Singleton bound can be achieved by a maximum distance separable code, as for example a Reed-Solomon code, such a code also suffices to achieve the capacity in the single-source single-sink case.

In the network multicast scenario, the situation is more complicated.
For the single-source multicast the capacity region was shown (\cite{RaymondNECC2002,YC06,CY06}) to be the same as (\ref{single-source-capacity}), with $C$ now representing the minimum of the min-cuts~\cite{YC06}. However, unlike single-source single-sink networks, in the case of single-source multicast, network error correction is required: network coding is required in general for multicast even  in the error-free case \cite{ACLY00}, and with the use of network coding errors  in the sink observations become dependent and cannot be corrected by end-to-end codes.

Two flavors of the network error correction problem are often considered. In the {\it coherent} case, it is assumed that there is centralized knowledge of the network topology and network code. Network error correction for this case was first addressed by the work of Cai and Yeung~\cite{RaymondNECC2002,YC06,CY06} for the single source scenario by generalizing classical coding theory to the network setting. However, their scheme has decoding complexity which is exponential in the network size.

In the harder {\it non-coherent} case, the network topology and/or network code are not known {\it a priori} to any of the honest parties. In this setting,~\cite{Jaggi_etal08, KK08} provided network error-correcting codes with a design and implementation complexity that is only polynomial in the size of network parameters. Reference~\cite{KK08} introduced an elegant approach where information transmission occurs via the space spanned by the received packets/vectors, hence any generating set for the same space is equivalent to the sink~\cite{KK08}. Error-correction techniques for this case were proposed in~\cite{KK08} and~\cite{SKK08} in the form of constant dimension and rank metric codes, respectively, where the codewords are defined as subspaces of some ambient space. These works considered only the single source case.

For the non-coherent multi-source multicast scenario \emph{without} errors, the scheme of ~\cite{Ho_etal06} achieves any point inside the rate-region. An extension of subspace codes to multiple sources, for a non-coherent multiple-access channel model without errors, was
provided in~\cite{SFD08}, which gave practical achievable  (but not rate-optimal) algebraic code constructions, and in~\cite{FragouliITW09MultiSource}, which derived the capacity region and gave a rate-optimal scheme for two sources.
For the multi-source case with errors,~\cite{FragouliMibiHoc09MultiSource} provided an efficient code construction achieving a strict subregion of the capacity region.

\section{Challenges}
\label{sec_challenges}

\begin{figure}
 \centerline{\includegraphics[scale=0.9]{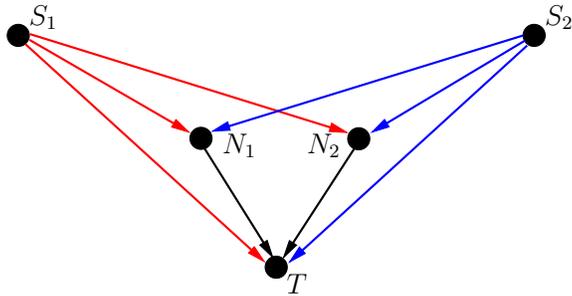}}
 \caption{A simple example to show that in the multiple source case
   in-network coding is required to achieve the network error
   correction capacity.}
 \label{fig:simple_example}
\end{figure}

In this work we address the capacity region and the corresponding code design for the multiple-source multicast communication problem under different adversarial scenarios. The issues which arise in this problem are best explained with a simple example for a single sink, which is shown in Fig.~\ref{fig:simple_example}. Suppose that the sources $\sources_1$ and $\sources_2$ encode their information independently from each other. We can allocate one part of the network to carry only information from $\sources_1$, and another part to carry only information from $\sources_2$. In this case only one source is able to communicate reliably under one link error. However, if coding at the middle nodes $N_1$ and $N_2$ is employed, the two sources are able to share network capacity to send redundant information, and each source is able to communicate reliably at capacity $1$ under a single link error. This shows that in contrast to the single source case, coding across multiple sources is required, so that sources can simultaneously use shared  network capacity to send redundant  information, even for a single sink.

In Section~\ref{The_Non_Coherent_Case} we show that for the example network in Fig.~\ref{fig:simple_example}, the capacity region is given by
\begin{align}
R_1 &\leq m_{\sources_1}-2z \nonumber\\
R_2 &\leq m_{\sources_2}-2z \label{eg:2source}\\
R_1+R_2 &\leq m_{\sources_1,\sources_2}-2z\nonumber,
\end{align}
where for $i=1,2$, rate $R_i$ is the information rate of $\sources_i$, min-cut $m_{\sources_i}$ is the minimum cut capacity between $\sources_i$ and sink $T$, min-cut $m_{\sources_1,\sources_2}$ is the minimum cut capacity between  $\sources_1$, $\sources_2$  and $T$ and $z$ is the known upper bound on the number of link errors. Hence, similarly to single-source multicast, the capacity region of a multi-source multicast network is described by the cut-set bounds. From that perspective, one may draw a parallel with point-to-point error-correction. However, for multi-source multicast networks point-to-point error-correcting codes do not suffice and a careful network code design is required. For instance, the work of~\cite{FragouliMibiHoc09MultiSource}, which applies single-source network error-correcting codes for this problem, achieves a rate-region that is strictly smaller than the capacity region  (\ref{eg:2source}) when $m_{\sources_1} +m_{\sources_2}\ne m_{\sources_1,\sources_2}$~\cite{fragouli_personal}.

\section{Our results}
\label{Our_results}

In this paper we consider a  ``side-channel'' model and an ``omniscient" adversarial model. In the former,
the adversary does not have access to all the information available in the network, for example as in~\cite{Jaggi_etal08, landberg08} where the sources share a secret with the sink(s) in advance of the network communication.  Let $\sources$ be the set of sources in the network, $\SourceNum$ be the number of sources, $R_i$ be the multicast transmission rate from source $\sources_i$, $1\leq i\leq \SourceNum$, to every sink, and for any non-empty subset $\sources'\subseteq\sources$ let $m_{\sources'}$ be the  minimum min-cut capacity between any sink and $\sources'$.

In Section~\ref{The_Random_Secret_Model} we prove the following theorem:
\begin{theorem}
\label{thm:shared_secret}
Consider a multiple-source multicast network error-correction problem on network $\graph$--possibly with unknown topology--where each source shares a random secret with each of the sinks. For any errors on up to $z$ links, the capacity region is given by:
\begin{align}
\displaystyle\sum_{i\in {\cal{I}}(\sources')} R_i \leq m_{\sources'} -z \ \forall\sources'\subseteq \sources.
\label{eq:thm_shared_secret}
\end{align}
and every point in the rate region can be achieved with a polynomial-time code.
\end{theorem}
\noindent By capacity region we mean the closure of all rate tuples $(R_1,\ldots,R_\SourceNum)$ for which there is a sequence of codes of length $\ell$, message sets $\mathcal{J}^i_\ell=\{1,\ldots,J^i_\ell\}$ and encoding and decoding functions $\{f_\ell^i\}, \{\phi_\ell^j\}$ for every node $i$ in the network and every sink $j$, so that for every $\epsilon>0$ and $\delta>0$ there is integer $L(\epsilon, \delta)>0$ such that for every $\ell>L(\epsilon, \delta)$ we have $\frac{1}{\ell}\log |\mathcal{J}^i|\geq R_i-\epsilon$ and the probability of decoding error at any sink is less than $\delta$ regardless of the message.

In ``omniscient" adversarial model, we do not assume any limitation on the adversary's knowledge, i.e.~decoding should succeed for {\it arbitrary} error values.  In
Section~\ref{General_approach} we derive the
multiple-access network error-correction capacity for both the coherent and non-coherent case. We show that network error-correction coding allows redundant network capacity to be shared among multiple sources, enabling the sources to simultaneously communicate reliably at their individual cut-set capacities under adversarial errors. Specifically, we prove the following theorem:
\begin{theorem}
\label{n_sources}
Consider a multiple-source multicast network error-correction problem on network $\graph$ whose topology may be unknown. For any errors on up to $z$ links, the capacity region is given by:
\begin{align}
\displaystyle\sum_{i\in {\cal{I}}(\sources')} R_i \leq m_{\sources'} -2z \ \forall\sources'\subseteq \sources.
\label{eq_non-coherent}
\end{align}
\end{theorem}
\noindent The rate-regions are, perhaps not surprisingly, larger for the
side-channel model than for the omniscient adversarial model.

Finally, in Section~\ref{Polynomial_time_construction} we provide computationally efficient distributed schemes for the non-coherent case (and therefore for the coherent case too) that are rate-optimal for correction of network errors injected by computationally unbounded adversaries. In particular, our code construction achieves decoding success probability at least
$1-|\SourceNum|\NetSize/p$
where $p$ is the size of the finite field $\F_p$ over which coding is performed, with complexity  $O(\ell m^{|\sources|})$, which is polynomial in the network size.

The remainder of the paper is organized as follows: In Section~\ref{Sec:Pre} we formally introduce our problem and give some mathematical preliminaries. In Section~\ref{The_Random_Secret_Model} we derive the capacity region and construct multi-source multicast error-correcting codes for the side-channel model. In Section~\ref{The_Non_Coherent_Case}, we consider two network error-correction schemes for omniscient adversary models which are able to achieve the full capacity region in both the coherent and non-coherent case. In particular, we provide a general approach based on minimum distance decoding, and then refine it to a practical code construction and decoding algorithm which has polynomial complexity (in all parameters except the number of sources). Furthermore, our codes are fully distributed in the sense that different sources require no knowledge of the data transmitted by their peers, and end-to-end, {\it i.e.} all nodes are oblivious to the adversaries present in the network and simply implement random linear network coding~\cite{RandCode0}. A remaining bottleneck is that while the implementation complexity (in terms of packet-length, field-size, and computational complexity) of our codes is polynomial in the size of most network parameters, it increases exponentially with the number of sources. Thus, the design of efficient schemes for a large number of sources is still open. Portions of this work were presented in~\cite{Svit_rate_regions} and in~\cite{hongyi_ted}.

\section{Preliminaries}
\label{Sec:Pre}

\subsection{Model}
\label{The_Model}

We consider a delay-free acyclic network $\Graph=(\Nodes,\Edges)$ where $\Nodes$ is the set of nodes and $\Edges$ is the set of edges. The capacity of each edge is normalized to be one symbol of the finite field $\F_p$ per unit time where $p$ is a power of a prime. Edges with non-unit capacity are modeled as parallel
edges.

There are two subsets $\sources, \sinks\subseteq\Nodes$ of nodes where $\sources=\{\sources_1,\sources_2, \ldots, \sources_\SourceNum\}$ is a set of $\SourceNum$ sources and $\sinks$ is a set of sinks within the network. Let $R_i$ be the multicast transmission rate from $\sources_i$, $1\leq i\leq \SourceNum$, to every sink. For any non-empty subset $\sources '\subseteq \sources$, let ${\cal{I}}(\sources')\subseteq\{1,2,\ldots,\SourceNum\}$ be the indices of the source nodes that belong to $\sources'$. Let $m_{\sources'}$ be the  minimum min-cut capacity between $\sources'$ and any sink. For each $i$, let ${\cal{C}}_i$ be the code used by source $i$. Let $\code_{\sources'}$ be the Cartesian product of the individual codes of the sources in $\sources '$.

Within the network there is a computationally unbounded adversary who can observe all the transmissions and inject its own packets on up to $z$ links\footnote{Note that since each transmitted symbol in the network is from a finite field, modifying symbol $x$ to symbol $y$ is equivalent to injecting/adding symbol $y-x$ into $x$.} that may be chosen as a function of his knowledge of the network, the message, and the communication scheme. The location of the $z$ adversarial links is fixed but unknown to the communicating parties. In case of a {\it{side-channel model}}, there additionally exists a random secret shared between all sources and each of the sinks as in~\cite{Jaggi_etal08, landberg08}.

The sources on the other hand do not have any knowledge about each other's transmitted information or about the links compromised by the adversary. Their goal is to judiciously add redundancy into their transmitted packets so that they can achieve any rate-tuple within the capacity region.

\subsection{Random Linear Network Coding}
\label{The_Random_Linear_Network_Coding}

In this paper, we consider the following well-known distributed random linear coding scheme~\cite{RandCode0}.

\emph{Sources:} All sources have incompressible data which they wish to
deliver to all the destinations over the network. Source $\sources_i$
arranges its data into batches of $b_i$ packets and insert these
packets into a $b_i \times \ell$ {\it message matrix} $M_i$ over
$\F_p$ (the \emph{packet-length} $\ell$ is a network design
parameter). Each source $\mathcal {S}_i$ then takes independent and
uniformly random linear combinations over $\F_p$ of the rows of $M_i$
to generate the packets transmitted on each outgoing edge.

\emph{Network nodes:} Each internal node similarly takes (uniformly) random linear combinations of the packets on its incoming edges to
generate packets transmitted on its outgoing edges.

\emph{Adversary:} The adversarial packets are defined as the difference between the received and transmitted packets on each link. They are similarly arranged into a matrix $Z$ of size $z\times \ell$.

\emph{Sink:} Each sink $\sink\in\sinks$ constructs a $B \times
\ell$ matrix $Y$ over $\F_p$ by treating the received packets as
consecutive length-$\ell$ row vectors of $Y$. Since all the operations in the network are linear, each sink has an incoming matrix $Y$ that is given by
\begin{align}
Y =T_1 M_1+ T_2M_2+\ldots+T_\SourceNum M_\SourceNum+T_z Z,
\label{eq:net_tran}
\end{align}
where $T_i$, $1\leq i\leq \SourceNum$, is the overall transform matrix
from $\sources_i$ to $\sink\in\sinks$ and $T_z$ is the overall
transform matrix from the adversary to sink $\sink\in\sinks$.

\subsection{Finite Field Extensions}
\label{The_Finite_Field_Extensions}

In the analysis below denote by $\F_p\hspace{-1.1mm}^{m\times n}$ the set of all $m\times n$ matrices with elements from $\F_p$. The identity matrix with dimension $m\times m$ is denoted by $\identity_m$, and the zero matrix of any dimension is denoted by $\zero$. The dimension of the zero matrix will be clear from the context stated. For clarity of notation, vectors are in bold-face ({\it e.g.} ${\bf A}$).

Every finite field $\F_p$, where $p$ can be \emph{algebraically extended}\footnote{Let $\F_p[x]$ be the set of all polynomials over $\F_p$ and $f(x)\in \F_p[x]$ be an irreducible polynomial of degree $n$. Then $\F_p[x]/f(x)$ defines an algebraic extension field $\F_{p^n}$ by a homomorphic mapping~\cite{Algebra_Martin}.}~\cite{Algebra_Martin} to a larger finite field $\F_q$, where $q=p^n$ for any positive integer $n$. Note that $\F_q$ includes $\F_{p}$ as a subfield; thus any matrix $A\in \F_p\hspace{-1.1mm}^{m\times\ell}$ is also a matrix in $\F_q\hspace{-1mm}^{m\times \ell}$. Hence throughout the paper, multiplication of matrices from different fields (one from the base field and the other from the extended field) is allowed and is computed over the extended field.

The above extension operation defines a bijective mapping between $\F_p\hspace{-1.1mm}^{m\times n}$ and $\F_{q}\hspace{-1mm}^{m}$ as follows:
\begin{itemize}
\item  For each $A\in\F_p\hspace{-1.1mm}^{m\times n}$, the folded version of $A$ is a vector $\mathbf{A}^f$
 in $\F_q\hspace{-1mm}^m$ given by $A \mathbf{a}^\text{T}$ where $\mathbf{a}=\{a_1,\ldots,a_n\}$ is a basis of the extension field $\F_q$ with respect to $\F_p$. Here we treat the $i^\text{th}$ row of $A$ as a single element in $\F_q$ to obtain the $i^\text{th}$ element of $\mathbf{A}^f$.

\item For each $\mathbf{B}\in\F_{q}\hspace{-1mm}^{m}$, the unfolded version of $\mathbf{B}$ is a matrix ${B}^u\in\F_p\hspace{-1.1mm}^{m\times n}$. Here we treat the $i^\text{th}$ element of $\mathbf{B}$ as a row in $\F_p\hspace{-1.1mm}^{1\times n}$ to obtain the $i^\text{th}$ row of ${B}^u$.
\end{itemize}

We can also extend these operations to include more general
scenarios. Specifically any matrix $A\in\F_p\hspace{-1.1mm}^{m\times
\ell n}$ can be written as a concatenation of matrices $A =
[A_1\ldots A_\ell]$, where $A_i\in F_p\hspace{-1.1mm}^{m\times n}$.
The folding operation is defined as follows: ${A}^f =
[\mathbf{A}_1^f\ldots \mathbf{A}_\ell^f]$. Similarly the unfolding
operation $u$ can be applied to a number of submatrices of a large
matrix, e.g., $[\mathbf{A}_1^f\ldots \mathbf{A}_\ell^f]^u=
[(\mathbf{A}_1^f)^u\ldots (\mathbf{A}_\ell^f)^u]=[A_1\ldots
A_\ell]$.

In this paper \emph{double algebraic extensions} are also considered. More precisely let $\F_Q$ be an algebraic extension from $\F_q$, where $Q=q^N=p^{nN}$ for any positive integer $N$. Table~\ref{Tab:Field-Notation} summarizes the notation of the fields
considered.

\begin{table}[ht]
\caption{Summary of field notations} 
\centering 
\begin{tabular}{|c|c|c|c|}
\hline 
Field&$\F_p$&$\F_q$&$\F_Q$
\\ [0.5ex]
\hline
Size &$p$&$q=p^n$&$Q=q^N $\\
\hline
\end{tabular}
\label{Tab:Field-Notation}
\end{table}

\noindent {\bf Note:} Of the three fields $\F_p$, $\F_{q}$ and
$\F_{Q}$ defined above, two or sometimes all three appear
simultaneously in the same equation. To avoid confusion, unless
otherwise specified, the superscript $f$ for folding is from  $\F_p$
to $\F_{q}$, and the superscript $u$ for unfolding is from $\F_{q}$
(or $\F_Q$) to $\F_p$.

\subsection{Subspace codes}
\label{The_Subspace_Codes}

In~\cite{KK08} an algebraic framework was developed for the non-coherent network scenario in the single-source case. The idea behind it is to treat the fixed-length packets as the vector subspaces spanned by them. Then what really matters at the decoder is the subspace spanned by the received packets rather than the individual packets.

Let $V$ be the vector space of length-$\ell$ vectors over the finite field $\F_p$, representing the set of all possible values of packets transmitted and received in the network. Let ${\cal{P}}(V)$ denote the set of all subspaces of $V$. A code $\cC$ consists of a nonempty subset of ${\cal{P}}(V)$, where each codeword $U\in\cC$ is a subspace of constant dimension.

Subspace errors are defined as additions of vectors to the transmitted subspace and subspace erasures are defined as deletions of vectors from the transmitted subspace.  Note that depending on the network code rate and network topology, network errors and erasures translate differently to subspace errors and erasures. For instance, subject to the position of adversary in the network, one network error can result in both dimension addition and deletion (i.e., both subspace error and subspace erasure in our terminology).  Let $\rho$ be the number of subspace erasures and let $t$ be the number of subspace errors caused by $z$ network errors.

The subspace metric~\cite{KK08} between two vector spaces $U_1,U_2\in {\cal{P}}(V)$ is defined as
\begin{align*}
d_{S}(U_1,U_2)&\doteq \dim(U_1+U_2) - \dim(U_1 \cap U_2)\\
              &= \dim(U_1)+\dim(U_2) - 2\dim(U_1 \cap U_2).
\end{align*}
In \cite{KK08} it shown that the minimum subspace distance decoder can successfully recover the transmitted subspace from the received subspace if
\begin{align*}
2(\rho+t) < D_{S}^\text{min},
\end{align*}
where $D_{S}^\text{min}$ is the minimum subspace distance of the code. Note that $d_S$ treats insertions and deletions of subspaces symmetrically. In \cite{sv09noncoherent} the converse of this statement for the case when information is transmitted at the maximum rate was shown.

In \cite{silvametric} a different metric on $V$, namely, the injection metric, was introduced and shown to improve upon the subspace distance metric for decoding of non-constant-dimension codes. The injection metric between two vector spaces $U_1,U_2\in{\cal{P}}(V)$ is defined as
\begin{align*}
d_{I}(U_1,U_2)&\doteq \max({\dim(U_1),\dim(U_2)}) -\dim(U_1 \cap U_2) \\
              &= \dim(U_1+U_2) - \min({\dim(U_1),\dim(U_2)}).
\end{align*}
$d_I$ can be interpreted as the number of error packets that an adversary needs to inject in order to transform input space $U_1$ into an output space $U_2$. The minimum injection distance decoder is designed to decode the received subspace as with as few error injections as possible. Note that for constant-dimensional codes $d_S$ and $d_I$ are related by
\begin{eqnarray*}
d_{I}(U_1,U_2)=\frac{1}{2}d_{S}(U_1,U_2).
\end{eqnarray*}

\subsection{Gabidulin Codes and Rank Metric Codes}
\label{sec:Gabidulin}

Gabidulin in~\cite{Gabidulin_TheoryOfCodes_1985} introduced a class
of error correcting codes over $\F_p\hspace{-1.1mm}^{m\times n}$.
Let $\mathbf{X}\in\F_{q}^{R}$ be the information vector,
$G\in\F_{q}^{m\times R}$ be the generator matrix,
$(G\mathbf{X})^u\in\F_p ^{m\times n}$ be the transmitted matrix,
$Z\in\F_p ^{m\times n}$ be the error matrix, and $(G
\mathbf{X})^u+Z\in\F_p ^{m\times n}$ be the received matrix. Then
decoding is possible if and only if
rank$(Z)\leq\lfloor\frac{d}{2}\rfloor$, where $d=m-R+1$ is the
minimum distance of the code.

The work of~\cite{SKK08} utilizes the results of~\cite{Gabidulin_TheoryOfCodes_1985} to obtain network error-correcting codes with the following properties:
\begin{theorem}[Theorem 11 in~\cite{SKK08}]
Let $Z$ be expressed as
$Z=\sum_{i\in[1,\tau]}\mathbf{L}_i\mathbf{E}_i$, such that:
\begin{itemize}
\item For each $i\in[1,\tau]$, $\mathbf{L}_i\in \F_p\hspace{-1.1mm}^{m\times 1}$ and $\mathbf{E}_i\in \F_p\hspace{-1.1mm}^{1\times n}$;

\item For each $i\in [1,\mu]$, $\mathbf{L}_i$ is known {\it a priori} by the sink;

\item For each $i\in [\mu+1,\mu+\delta]$, $\mathbf{E}_i$ is known {\it a priori} by the sink;

\item $2\tau-\mu-\delta\leq d-1$,
\end{itemize}
using Gabidulin codes the sink can decode $\mathbf{X}$ with at
most $\mathcal O(mn)$ operations over $\F_{q}$.
\label{Th:GabiDecode}
\end{theorem}
When $\mu=\delta=0$, Theorem~\ref{Th:GabiDecode} reduces to the
basic case where the sink has no prior knowledge about
$Z$.

For any matrices $B_1\in \F_p\hspace{-1.1mm}^{m_1\times m}$ and
$B_2\in \F_p\hspace{-1.1mm}^{m_2\times m}$ the following proposition holds and is a direct consequence of Corollary $3$ in~\cite{SKK08}:
\begin{proposition}
$d_{S}(\langle B_1\rangle,\langle B_2\rangle)\leq 2 \text{rank}(B_1-B_2)$
\label{pro:rowdis-rankdis}
\end{proposition}
where $\langle B_1\rangle$, $\langle B_2\rangle$ are the row-spaces of matrices $B_1, B_2$ respectively.

\section{Side-Channel Model}
\label{The_Random_Secret_Model}

The side-channel model is an extension of the random secret model
considered in~\cite{landberg08} to the case of multiple sources. In
that model every source shares a uniformly distributed random secret
with each of the sinks. For each source the ``secret'' consists of a
set of symbols drawn uniformly at random from the base field $\F_p$
and the adversary does not have access to these secret symbols. This
set of uniformly random symbols can be shared between each source and
the sinks either before the transmission starts or during the
transmission through a low capacity channel that is secret from the
adversary and cannot be attacked by it. Each source has a different
secret from all the other sources which makes this scheme distributed.
\begin{IEEEproof}[Proof of Theorem~\ref{thm:shared_secret}]
\emph{Converse:} Let $l_{i,j},j = 1,\dots,n_i,$ be the outgoing links of each source $S_i,i = 1, \ldots, s$. Take any $\sources'\subseteq \sources$. We construct the graph $\graph_{\sources'}$ from $\graph$ by adding  a virtual super source node $w_{\sources'}$, and $n_i$ links $l'_{i,j},j = 1,\dots,n_i,$ from $w_{\sources'}$ to source $S_i$ for each $i \in {\cal{I}}(\sources')$. Note that the minimum cut capacity between $w_{\sources'}$ and any sink is at least $m_{\sources'}$. Any network code that multicasts rate $R_i$ from each source $S_i, i \in {\cal{I}}(\sources')$ over $\graph$ corresponds to a network code that multicasts rate $\displaystyle\sum_{i\in {\cal{I}}(\sources')} R_i$ from $w_{\sources'}$  to all sinks over $\graph_{\sources'}$; the symbol on each link $l'_{i,j}$ is the same as that on link $l_{i,j}$, and the coding operations at all other nodes are identical for $\graph$ and $\graph_{\sources'}$. For the case of a single source, the adversary can choose the $z$ links on the min-cut and set their outputs equal to zero. Therefore in this case the maximum possible achievable rate $R$ is
\begin{align}
R\leq C-z
\label{eqn:C_minus_z}
\end{align}
where $C$ is the multicast min-cut capacity of the network. The converse follows from applying inequality (\ref{eqn:C_minus_z}) to $w_{\sources'}$ for each $\sources'\subseteq \sources$.
\end{IEEEproof}

\emph{Achievability:} In the case of the side-channel model, for notational convenience, we will restrict ourselves to the analysis of the situation where there are only two sources $\sources_1,\sources_2\in \Nodes$ transmitting information to one sink $\sink\in \Nodes$, since the extension of our result to more sources and sinks is straightforward and analyzed briefly in Section~\ref{more_than_two_sources}.

\noindent\emph{Encoding:} Source $\sources_1$ encodes its data into matrix $X_1\in\F_p\hspace{-1.1mm}^{R_1\times(\ell-\alpha)}$ of size $R_1\times(\ell-\alpha)$, where $\alpha=m_{\sources_1,\sources_2}^2+1$, with symbols from $\F_p$ and arranges its message into $M_1=\begin{bmatrix}L_1&X_1\end{bmatrix}$ where $L_1\in\F_p\hspace{-1.1mm}^{R_1\times\alpha}$ is a matrix that will be defined below. Similarly, source $\sources_2$ arranges its data into matrix $M_2=\begin{bmatrix}L_2&X_2\end{bmatrix}$ where $L_2\in\F_p\hspace{-1.1mm}^{R_2\times\alpha}$ will be defined below and $X_2\in\F_p\hspace{-1.1mm}^{R_2\times(\ell-\alpha)}$.

The shared secret between source $\sources_i$ and sink $\sink$ is
composed of a length--$\alpha$ vector $W_i=\begin{bmatrix}w_{i1}&\ldots&w_{i\alpha}\end{bmatrix} \in\F_p\hspace{-1.1mm}^{1\times\alpha}$ and a matrix $H_i\in\F_p\hspace{-1.1mm}^{R_i\times\alpha}$, where the
elements of both  $W_i$ and $H_i$ are drawn uniformly at random from
$\F_p$. The vector $W_i$ defines a \emph{parity-check} matrix
$P_i\in\F_p\hspace{-1.1mm}^{\ell\times\alpha}$ whose $(m,n)$-th entry
equals $\left(w_{in}\right)^m$, {\it i.e.}, the element $w_{in}$ taken
to the $m^\text{th}$ power. The matrix $L_i$ is defined so that the following equality holds
\begin{eqnarray}
H_i &=& M_i P_i
= \begin{bmatrix}L_i&X_i\end{bmatrix} \begin{bmatrix}V_i\\--\\\tilde{P}_i\end{bmatrix}
= L_i V_i+X_i \tilde{P}_i
\label{eqn:Mi_times_Pi}
\end{eqnarray}
where $V_i$, $\tilde{P}_i$ correspond to rows $\{1,\ldots,\alpha\}$ and $\{\alpha+1,\ldots,\ell\}$ of matrix $P_i$ respectively. Matrix $V_i\in\F_p\hspace{-1.1mm}^{\alpha\times\alpha}$ is a Vandermonde matrix and is invertible whenever vector $W_i$ contains pairwise different non-zero elements from $\F_p$, else $W_i$ is non-invertible which happens with probability at most $\alpha^2/p$ (each of the elements $w_{ij}$ is zero or identical to another element with probability at most $\alpha/p$). Whenever the matrix $V_i$ is invertible source $\sources_i$ solves equation (\ref{eqn:Mi_times_Pi}) to find $L_i$ and substitutes it into matrix $M_i$. When the matrix $V_i$ is non-invertible then $L_i$ is substituted with the zero matrix.

\noindent\emph{Linear Coding:} Once matrices $M_1$, $M_2$ are formed then both sources and the internal nodes perform random linear network coding operations and therefore sink $\sink$ gets
\begin{align}
&Y = T_1 M_1+ T_2M_2+T_z Z\notag\\
\Leftrightarrow &Y = \begin{bmatrix}T_1&T_2&T_z\end{bmatrix} \begin{bmatrix}M_1\\--\\M_2\\--\\Z\end{bmatrix}
\label{eq:net_tran_two_sources}
\end{align}
where $T_i\in\F_p\hspace{-1.1mm}^{m_{\sources_1,\sources_2}\times R_i}$ and $T_z\in\F_p\hspace{-1.1mm}^{m_{\sources_1,\sources_2}\times z}$.

\noindent\emph{Decoding:} Assume that matrix $Y\in\F_p\hspace{-1.1mm}^{m_{\sources_1,\sources_2}\times \ell}$ has column rank equal to $r$ and matrix $Y^s\in\F_p\hspace{-1.1mm}^{m_{\sources_1,\sources_2}\times r}$ contains $r$ linearly independent columns of $Y$. Since all the columns of $Y$ can be written as linear combinations of columns of $Y^s$, then $Y = Y^sF$ where $F\in\F_p\hspace{-1.1mm}^{r\times \ell}$. The columns of $M_1$, $M_2$ and $Z$ corresponding to those in $Y^s$ are denoted as $M_1^s\in\F_p\hspace{-1.1mm}^{R_1\times r}$, $M_2^s\in\F_p\hspace{-1.1mm}^{R_2\times r}$ and $Z^s\in\F_p\hspace{-1.1mm}^{z\times r}$ respectively. Therefore
\begin{align}
Y^s = \begin{bmatrix}T_1&T_2&T_z\end{bmatrix} \begin{bmatrix}M_1^s\\--\\M_2^s\\--\\Z^s\end{bmatrix}
\label{eq:net_tran_two_sources_s}
\end{align}
and by using equations (\ref{eq:net_tran_two_sources}), (\ref{eq:net_tran_two_sources_s}) we have
\begin{align*}
Y=Y^sF\displaystyle\mathop{\Rightarrow}^{(\ref{eq:net_tran_two_sources})}_{(\ref{eq:net_tran_two_sources_s})} \begin{bmatrix}T_1&T_2&T_z\end{bmatrix} \begin{bmatrix}M_1\\--\\M_2\\--\\Z\end{bmatrix}=\begin{bmatrix}T_1&T_2&T_z\end{bmatrix} \begin{bmatrix}M_1^s\\--\\M_2^s\\--\\Z^s\end{bmatrix}F.
\end{align*}
Therefore $M_1=M_1^sF$ and $M_2=M_2^sF$ since for large enough $p$, matrix $\begin{bmatrix}T_1&T_2&T_z\end{bmatrix}$ is invertible with high probability~\cite{Jaggi_etal08}. Consequently, equation (\ref{eqn:Mi_times_Pi}) can be written as $M_1^s(FP_1)=H_1$ where matrices $F$, $P_1$ and $H_1$ are known and matrix $M_1^s$ is unknown and can be found using standard Gaussian elimination.

As in~\cite{Jaggi_etal08} it can be proved that the solution obtained by the Gaussian elimination is with high probability the unique solution to equation $M_1P_1=H_1$. Indeed, using Claim 5 of~\cite{Jaggi_etal08}, for any $\hat{M}_1^s\neq M_1^s$ the probability (over $w_{11},\ldots,w_{1\alpha}$) that $\hat{M}_1^s(FP_1)=H_1$ is at most $\left(\frac{\ell}{p}\right)^\alpha$. Since there are $p^{R_1\cdot r}$ different matrices $\hat{M}_1$ $(\hat{M}_1=\hat{M}_1^sF$ and $\hat{M}_1^s\in\F_p\hspace{-1.1mm}^{R_1\times r})$ by taking the union bound over all different $\hat{M}_1$ (Corollary $6$ in~\cite{Jaggi_etal08}) we conclude that the probability of having more than one solution for equation $M_1P_1=H_1$ is at most $p^{R_1\cdot m_{\sources_1,\sources_2}}\left(\frac{\ell}{p}\right)^\alpha<\frac{\ell^\alpha}{p}$. Decoding of $X_2$ is similar.

\noindent\emph{Probability of error analysis:} In order for the decoding to fail one or more of the following three events should occur:
\begin{enumerate}
\item At least one of the network transform matrices $\begin{bmatrix}T_1&T_2&T_z\end{bmatrix}$ is not full column rank. According to~\cite{RandCode0}, this happens with probability less than {\footnotesize$\nchoosek{|\Edges|}{z}$}$\frac{|\Edges||\sinks|}{p}$, where $|\Edges|$, $|\sinks|$ is the number of edges and the number of sinks in the network. Term {\footnotesize${\scriptstyle\nchoosek{|\Edges|}{z}}$} is the number of different sets of $z$ links the adversary can attack and $\frac{|\Edges|}{p}$ is an upper bound for the probability that matrix $\begin{bmatrix}T_1&T_2&T_z\end{bmatrix}$ is not full column rank when the adversary has attacked a specific set of links.
\item Either of the Vandermonde matrices $V_1$ or $V_2$ are not invertible. By using the union bound this happens with probability at most $2\alpha^2/p$.
\item There are more than one solutions for equations $\hat{M}_i^s(FP_i)=H_i$ for $i\in\{1,2\}$. This happens with probability at most $2\ell^\alpha/p=2\ell^{(m_{\sources_1,\sources_2}^2+1)}/p$.
\end{enumerate}

\footnote{From the three probability events the third one dominates the other two when packet size is large.}Hence, it is not difficult to see that the probability of decoding failure can be made arbitrarily small as the size $p$ of the finite field increases. Moreover increasing $\ell$ without bound we can approach any point inside the rate-region. The decoding complexity of the algorithm is dominated by the complexity of the Gaussian elimination that is $O(\ell m_{\sources_1,\sources_2}^3)$.

\section{Omniscient Adversarial Model}
\label{The_Non_Coherent_Case}
\subsection{General approach}
\label{General_approach}

In this section we construct capacity-achieving codes for the multiple-source multicast non-coherent network scenario. We use the algebraic framework of subspace codes developed in~\cite{KK08}, which provides a useful tool for network error and erasure correction over general unknown networks. In Section~\ref{The_Subspace_Codes}, we gave basic concepts and definitions of subspace network codes needed for further discussion.

In the proof of Theorem~\ref{n_sources} we show how to design non-coherent network codes that achieve upper bounds given by (\ref{eq_non-coherent}) when a  minimum (or bounded) injection distance decoder is used at the sink nodes. Our code construction uses random linear network coding at intermediate nodes, single-source network error-correction capacity-achieving codes at each source, and an overall global coding vector.
Our choice of decoder relies on the observation that subspace erasures are not arbitrarily chosen by the adversary, but also depend on the network code.  Since, as we show below, with high probability in a random linear network code, subspace erasures do not cause confusion between transmitted codewords, the decoder focuses on the discrepancy between the sent and the received codewords caused by subspace errors. The error analysis shows that injection distance decoding succeeds with high probability over the random network code. On the other hand, the subspace minimum distance of the code is insufficient to account for the total number of subspace errors and erasures that can occur.  This is in contrast to constant dimension single-source codes, where subspace distance decoding is  equivalent to injection distance decoding~\cite{silvametric}.
\begin{IEEEproof}[Proof of Theorem~\ref{n_sources}]
\emph{Converse:} The proof is similar to the converse of the proof of Theorem~\ref{thm:shared_secret} with the  exception that after connecting any subset of sources $\sources'\subseteq \sources$ by a virtual super-source node $w_{\sources'}$, we apply the network Singleton bound~\cite{YC06} to $w_{\sources'}$ for each $\sources'\subseteq \sources$.

\emph{Achievability:} 1) \emph{Code construction:} Consider any rate vector $(R_1, \ldots, R_\SourceNum)$ such that
\begin{align}
\displaystyle\sum_{i\in {\cal{I}}(\sources')} R_i < m_{\sources'} -2z \ \forall\sources'\subseteq \sources.
\label{star_eq}
\end{align}
Let each ${\mathcal{C}}_i$, $i=1, \ldots, \SourceNum$ be a code consisting of codewords that are $k_i-$dimensional linear subspaces.  The codeword transmitted by source $\sources_i$ is spanned by the packets transmitted by $\sources_i$. From the single source case, for each source $i=1,\ldots,\SourceNum$ we can construct a code $\cC_i$ where
\begin{align}
k_i>R_i+z
\label{equation:k}
\end{align}
that corrects any $z$ additions~\cite{Jaggi_etal08}. This implies that by \cite{sv09noncoherent}, $\cC_i$ has minimum subspace distance greater than $2z$, i.e. for any pair of distinct codewords $V_i$, $V'_i\in\cC_i$
\begin{eqnarray*}
d_S(V_i, V'_i)=\dim(V_i)+\dim(V'_i) - 2\dim(V_i\cap V'_i)>2z.
\end{eqnarray*}
Hence,
\begin{align}
\dim(V_i\cap V'_i) <k_i-z \quad \forall\ V_i,V'_i\in\cC_i.
\label{equation:vv}
\end{align}
By (\ref{equation:k}), we have:
\begin{eqnarray*}
\displaystyle\sum_{i \in {\cal{I}(\sources')}} k_i> \displaystyle\sum_{i \in {\cal{I}(\sources')}} R_i+ |\sources'|z.
\end{eqnarray*}
Therefore, by combining it with (\ref{star_eq}) and scaling all source rates and link capacities by a sufficiently large integer if necessary, we can assume without loss of generality that we can choose $k_i$ satisfying
\begin{align}
\displaystyle\sum_{i \in {\cal{I}(\sources')}} k_i\less m_{\sources'}+(|\sources'| -2)z\ \forall\sources'\subseteq \sources.
\label{equation:cuts}
\end{align}

We can make vectors from one source linearly independent of vectors from all other sources by prepending a length--$(\displaystyle\sum_{i\in
{\cal{I}}(\sources)} k_i)$ global encoding vector, where the $j$th global encoding vector, $j = 1, 2,\dots, \sum_{i\in
{\cal{I}}(\sources)} k_i$, is the unit vector  with a single nonzero entry in the $j$th position. This adds an overhead that becomes
asymptotically negligible as packet length grows. This ensures that
\begin{align}
\dim(V_i\cap V_j)=0 \ \forall i \ne j, V_i \in \cC_i, V_j \in \cC_j.
\label{intersection}
\end{align}
\emph{Error analysis:} Let $X \in {\cC_{\sources}}$ be the sent codeword, and let $R$ be the subspace received at a sink. Consider any $\sources'\subseteq\sources$. Let $\overline{\sources'}=\sources \setminus \sources'$. Let $X=V\oplus W$, where $V\in\code_{\sources'},W \in \cC_{\overline\sources'}$ and $V$ is spanned by the codeword $V_i$ from each code $\cC_i, i\in{\cal{I}}(\sources')$. We will show that with high probability over the random network code, there does not exist another codeword $Y=V' \oplus W$, such that $V'$ is spanned by a codeword $V_i' \neq V_i$ from each code $\cC_i, i\in{\cal{I}}(\sources')$, which could also have produced $R$ under arbitrary errors on up to $z$ links in the network.

Fix any sink $\sink$. Let $\cR$ be the set of packets (vectors) received by $\sink$, i.e. $R$ is the subspace spanned by $\cR$. Each of the packets in $\cR$ is a linear combination of vectors from $V$ and $W$ and error vectors, and can be expressed as $\bp =\bu_\bp+\bw_\bp$, where $\bw_\bp$ is in $W$ and the global encoding vector of $\bu_\bp$ has zero entries in the positions corresponding to sources in set ${\cal{I}}(\overline{\sources'})$.

The key idea behind our error analysis is to show that with high probability subspace deletions do not cause confusion, and that more than $z$ additions are needed for $X$ be decoded wrongly at the sink, i.e we will show that
\begin{align*}
 d_{I}(R,V' \oplus W)=\dim(R) - \dim(R \cap (V' \oplus W)) > z.
\end{align*}

Let $P =\mbox{span}\{\bu_\bp:\;\bp\in\cR\}$. Let $M$ be the matrix whose rows are the vectors $\bp\in\cR$, where the $j$th row of $M$ corresponds to the $j$th vector $\bp\in\cR$. Similarly, let $M_{\bu}$ be the matrix whose $j$th row is the vector $\bu_\bp$ corresponding to the $j$th vector $\bp\in\cR$, and let $M_{\bw}$ be the matrix whose $j$th row is the vector $\bw_\bp$ corresponding to the $j$th vector $\bp\in\cR$. Consider matrices $ A,B$ such that the rows of $AM_\bu$ form a basis for $P\cap V'$ and, together with the rows of $BM_\bu$,
form a basis for $P$. The linear independence of the rows of $\left[\begin{array}{c}AM_\bu\\BM_\bu\end{array}\right]$ implies that the rows of $\left[\begin{array}{c}AM\\BM\end{array}\right]$ are also linearly independent, since otherwise there would be a  nonzero matrix $D$ such that
\begin{align*}D\left[\begin{array}{c}AM\\BM\end{array}\right] = 0
&\Rightarrow D\left[\begin{array}{c}AM_\bw\\BM_\bw\end{array}\right]= 0\\&\Rightarrow D\left[\begin{array}{c}AM_\bu\\BM_\bu\end{array}\right]= 0,
\end{align*}a contradiction.
For $\bw_\bp$  in $W$, $\bu_\bp+\bw_\bp$ is in $V'\oplus W$ only if
$\bu_\bp$ is in $V'$, because the former implies $\bu_\bp = \bu_\bp+\bw_\bp-\bw_\bp$ is in $V'
\oplus W$ and since $\bu_\bp$ has zero entries in the positions of the global encoding vector corresponding to ${\cal{I}}(\overline{\sources'})$ it must be in $V'$.
Thus, since any vector in the row space of $BM_\bu$ is not in $V'$,  any vector in the row space of $BM$ is not in $V' \oplus W$.
Since the row space of $BM$ is a subspace of $R$, it follows that the number
of rows of $B$ is equal to $\dim(P) -\dim(P\cap V')$ and is less than or equal to $\dim(R) - \dim(R \cap (V' \oplus W))$.  Therefore,
\begin{align} \label{eqngoal}
d_{I}(R,V' \oplus W)&= \dim(R) - \dim(R \cap (V' \oplus W))\\
&\geq \dim(P) -\dim(P\cap V').\notag
\end{align}

We next show that for random linear coding in a sufficiently large field, with high probability
\begin{align}
\dim(P) -\dim(P\cap V') > z
\label{equation:difference}
\end{align}
for all $V'$ spanned by a codeword $V_i' \neq V_i$ from each code $\cC_i, i\in{\cal{I}}(\sources')$.

Consider first the network with each source $i$ in $\sources'$
transmitting $k_i$ linearly independent packets from $V_i$, sources
in $\overline{\sources'}$ silent, and no errors. From the
maxflow-mincut bound, any rate vector $(h_1,\ldots, h_{|\sources'|})$, such
that
\begin{align*}
\displaystyle\sum_{i\in \sources''} h_i \leq m_{\sources''}\quad \forall
\sources''\subseteq \sources'
\end{align*}
can be achieved. Combining this with (\ref{equation:cuts}), we can see that in the error-free case, each $s_i \in \sources'$ can transmit information to the sink at rate $k_i-\frac{(|\sources'| -2)z}{|\sources'|}$ for a total rate of
\begin{align}
\displaystyle\sum_{i\in{\cal{I}}(\sources')}k_i-(|\sources'| -2)z.
\label{total_rate}
\end{align}
With sources in $\overline{\sources'}$ still silent, consider the addition of $z$ unit-rate sources corresponding to the error links. The space spanned by the received packets corresponds to $P$.
Consider any $V'$ spanned by a codeword $V_i' \neq V_i$ from each code $\cC_i, i\in{\cal{I}}(\sources')$.

Let $Z$ be the space spanned by the error packets, and let $z' \less z$ be the minimum cut between the error sources and the sink. Let   $P=P_V\oplus P_Z$, where $P_Z=P\cap Z$ and $P_V$ is a subspace of $V$. There exists a routing solution, which we distinguish by adding tildes in our notation, such that $\dim\tilde{P}_Z=z'$ and, from (\ref{total_rate}), $\dim\tilde{P}\greater\displaystyle\sum_{i\in{\cal{I}}(\sources')}k_i-(|\sources'| -2)z$, so
\begin{align}
\dim(\tilde{P}_V)\greater\displaystyle\sum_{i\in{\cal{I}}(\sources')}k_i-(|\sources'|
-2)z-z'.
\label{two_stars}
\end{align}
Note that, by (\ref{intersection}), a packet from $V_i$ is not in any $V'_j\in\cC_j,j\ne i$, and hence is in $V'$ if and only if it is in $V'_i$. Therefore, by (\ref{equation:vv})
\begin{align*}
\dim(\tilde{P}_V \cap V') \leq
\displaystyle\sum_{i\in {\cal{I}}(\sources')} \dim (V_i \cap V_i') <
\displaystyle\sum_{i \in {\cal{I}}(\sources')}k_i-|\sources'|z.
\end{align*}
Therefore, using (\ref{two_stars}) we have
\begin{align*}
\dim(\tilde{P}_V\cup V') &=\dim(\tilde{P}_V) +\dim(V')-\dim(\tilde{P}_V\cap V')\\
&>\dim(\tilde{P}_V) +\dim(V') +|\sources'|z -\displaystyle\sum_{i \in {\cal{I}}(\sources')}k_i\\
&\geq \sum_{i \in {\cal{I}}(\sources')} k_i-(|\sources'| -2)z-z' +|\sources'|z \\
&=\sum_{i \in {\cal{I}}(\sources')} k_i +2z -z'\greater\sum_{i \in {\cal{I}}(\sources')} k_i +z.
\end{align*}
Then
\begin{align*}
  \dim(\tilde{P}\cup V') >\sum_{i \in {\cal{I}}(\sources')} k_i +z.
\end{align*}
For random linear coding in a sufficiently large field, with high probability by its generic nature
\begin{align*}
\dim(P\cup V')\greater\dim(\tilde{P}\cup V')>\displaystyle\sum_{i \in {\cal{I}}(\sources')} k_i +z,
\end{align*}
and this also holds for any $z$ or fewer errors, all sinks, and all  $V'$ spanned by a codeword $V_i' \neq V_i$ from each code $\cC_i, i\in{\cal{I}}(\sources')$. Then, (\ref{equation:difference}) follows by
\begin{align*}
\dim(P) -\dim(P\cap V') =\dim(P\cup V')-\dim(V').
\end{align*}

Hence, using (\ref{equation:difference}) and (\ref{eqngoal}),
\begin{align*}
d_{I}(R,V'\oplus W)&=\dim(R) -\dim(R\cap(V'\oplus W))\\
&\geq\dim(P) -\dim(P\cap V')>z.
\end{align*}
Thus, more than $z$ additions are needed to produce $R$ from $Y=V'\oplus W$. By the generic nature of random linear coding, with high probability this holds for any $\sources'$. Therefore, at every sink the minimum injection distance decoding succeeds with high probability over the random network code.\\
\emph{Decoding complexity:} Take any achievable rate vector $(R_1,
R_2, \ldots, R_s)$. For each $i=1, \ldots, s$, $S_i$ can transmit at
most $p^{R_i\ell}$ independent symbols. Decoding can be done by exhaustive search, where the decoder checks each possible set of
codewords to find the one with minimum distance from the observed set
of packets, therefore, the decoding complexity of the minimum
injection distance decoder is upper bounded by  $O(p^{l{\sum_{i=1}^s R_i}})$.
\end{IEEEproof}

\subsection{Polynomial-time construction}
\label{Polynomial_time_construction}

Similar to the side-channel model, we will describe the code for the case where there are only two sources $\sources_1,\sources_2\in \Nodes$ transmitting information to
one sink $\sink\in \Nodes$, since the extension of our
results to more sources and sinks is straightforward and analyzed briefly in Section~\ref{more_than_two_sources}. To further simplify the discussion we show the code construction for
rate-tuple $(R_1,R_2)$ satisfying $R_1\leq m_{\sources_1}-2z$, $R_2\leq m_{\sources_2}-2z$, $R_1+R_2+2z=m_{\sources_1,\sources_2}$ and exactly $m_{\sources_1,\sources_2}$ edges incident to sink $\sink$ (if more do, redundant information can be discarded).

\noindent\emph{Encoding:} Each source $\sources_i$, $i\in\{1,2\}$, organizes its information into a matrix $X_i\in\F_p\hspace{-1.1mm}^{R_i\times knN}$ with elements from $\F_p$, where $n=R_1+2z$, $N=R_2+2z$ and $k$ is an integer (and a network parameter). In order to correct adversarial errors, redundancy is introduced through the use of Gabidulin codes (see Section~\ref{sec:Gabidulin} for details).

More precisely the information of $\sources_1$ can be viewed as a matrix $X_1\in\F_q\hspace{-1mm}^{R_1\times kN}$, where $\F_q$ is an algebraic extension of $\F_p$ and $q = p^n$ (see Section~\ref{The_Finite_Field_Extensions} for details). Before transmission $X_1$ is multiplied with a generator matrix, $G_1\in\F_q\hspace{-1mm}^{n\times R_1}$, creating $G_1 X_1\in\F_q\hspace{-1mm}^{n\times kN}$ whose unfolded version $M'_1=(G_1 {X}_1)^u$ is a matrix in $\F_p\hspace{-1.1mm}^{n\times knN}$. The information of $\sources_2$ can be viewed as a matrix
$X_2\in\F_Q^{R_2\times k}$, where $\F_Q$ is an algebraic
extension of $\F_q$ where $Q=q^N=p^{nN}$. Before transmission
$X_2$ is multiplied with a generator matrix, $G_2\in\F_Q^{N\times R_2}$, creating $G_2 X_2\in\F_Q^{N\times k}$ whose unfolded version $M'_2=(G_2 X_2)^u$ over $\F_p$ is a matrix in $\F_p\hspace{-1.1mm}^{N\times knN}$. Both $G_1$ and $G_2$ are chosen as generator matrices for Gabidulin codes and have the capability of correcting errors of rank at most $z$ over $\F_p$ and $\F_q$ respectively.

In the scenario where sink $\sink$ does not know $T_1$ and $T_2$ {\it a priori} the two sources append  headers on their transmitted packets to convey information about $T_1$ and $T_2$ to the sink. Thus source $\sources_1$ constructs message matrix $M_1=\begin{bmatrix}\identity_n&\zero&M'_1\end{bmatrix}$ with the zero matrix $\zero$ having dimensions $n\times N$, and source $\sources_2$ constructs a message matrix $\begin{bmatrix}\zero&\identity_N&M'_2\end{bmatrix}$ with the zero matrix $\zero$ having dimension $N\times n$. Each row of matrices $M_1$, $M_2$ is a packet of length $\ell = knN+n+N$.

Before we continue with the decoding we need to prove the following two Lemmas:
\begin{lemma}
\label{lemma:RankBoundE}
Folding a matrix does not increase its rank.
\end{lemma}
\begin{IEEEproof}
Let  matrix $H\in\F_p\hspace{-1.1mm}^{m\times k n}$ has rank$(H)=r$ in field $\F_p$. Thus $H=W Z$, where $Z\in\F_p\hspace{-1mm}^{r\times k n}$ is of full row rank and $W\in\F_p\hspace{-1mm}^{m\times r}$ is of full column rank. After the folding operation $H$ becomes $H^f = W Z^f$ and therefore has rank in the extension field $\F_q$, where $q=p^n$, is at most $r$, {\it i.e.} rank$({H}^f)\leq r$.
\end{IEEEproof}

\begin{lemma}
Matrix $\begin{bmatrix}T_1G_1& T_2\end{bmatrix}\in\F_q\hspace{-1mm}^{m_{\sources_1, \sources_2}\times m_{\sources_1, \sources_2}}$ is
invertible with probability at least $1-\NetSize/p$.
\label{lemma:Invertible_D}
\end{lemma}
\begin{IEEEproof}
Let $\mathcal{X}$  be the set of random variables over $\F_p$ comprised of the local coding coefficients used in the random linear network code. Thus the determinant of $\begin{bmatrix}T_1G_1& T_2\end{bmatrix}$ is a polynomial $\textbf{f}(\mathcal{X})$ over $\F_q$ of degree at most $\NetSize$ (see Theorem 1 in~\cite{RandCode0} for details). Since the variables $\mathcal{X}$ in $\textbf{f}(\mathcal{X})$ are evaluated over $\F_p$, $\textbf{f}(\mathcal{X})$ is equivalent to a vector of polynomials $(f_1(\mathcal{X}),f_2(\mathcal{X}),\ldots, f_n(\mathcal{X}))$, where $f_i(\mathcal{X})\in \F_p[\mathcal{X}]$ is a polynomial over $\F_p$ with variables in $\mathcal{X}$. Note that $f_i(\mathcal{X})$ also has degree no more than $\NetSize$ for each $i\in\{1,\ldots,n\}$. Thus once we prove that there exists an evaluation of $\mathcal{X}$ such that $\textbf{f}$ is a nonzero vector over $\F_p$, we can show that matrix $\begin{bmatrix}T_1G_1& T_2\end{bmatrix}$ is invertible with probability at least $1-\NetSize/p$ by the Schwartz-Zippel lemma~\cite{CCBook} (Proposition 98).

Since $R_1+N=m_{\sources_1,\sources_2}$, $R_1\leq m_{\sources_1}$ and $N\leq m_{\sources_2}$, there exist $R_1+N$ edge-disjoint-paths:$\DisPath^1_1,\DisPath^1_2,\ldots,\DisPath^1_{R_1}$ from $\sources_1$ to $\sink$ and $\DisPath^2_1,\DisPath^2_2,\ldots,\DisPath^2_{N}$ from $\sources_2$ to $\sink$. The variables in $\mathcal{X}$ are evaluated in the following manner:
\begin{enumerate}
\item Let $\zero$ be the zero matrix in $F_q\hspace{-1mm}^{n \times N}$. We choose the variables in $\mathcal{X}$ so that the $R_1$ independent rows of $\begin{bmatrix}G_1 & \zero\end{bmatrix} \in\F_q\hspace{-1mm}^{n\times m_{\sources_1,\sources_2}}$ correspond to routing information from $\sources_1$ to $\sink$ via $\DisPath_1^1, \ldots, \DisPath_{R_1}^1$.

\item Let $\{\textbf{u}_{R_1+1},\textbf{u}_{R_1+2},\ldots,\textbf{u}_{m_{\sources_1,\sources_2}}\}$ be $N$ distinct rows of the identity matrix in $\F_q\hspace{-1mm}^{m_{\sources_1,\sources_2}\times m_{\sources_1,\sources_2}}$ such that for each $i\in\{1,\ldots,N\}$, $\textbf{u}_{R_1+i}$ has the element $1$ located at position $R_1+i$. Then these $N$ vectors correspond to routing information from $\sources_2$ to sink $\sink$ via $\DisPath^2_1,\DisPath^2_2,\ldots,\DisPath^2_N$.
\end{enumerate}

Under such evaluations of the variables in $\mathcal{X}$, matrix $\begin{bmatrix}T_1G_1& T_2\end{bmatrix}$ equals $\begin{bmatrix} G_1'&\zero\\\zero &\identity_N\end{bmatrix}$, where $G_1'\in \F_q\hspace{-1mm}^{R_1\times R_1}$ consists  of the $R_1$ independent rows of $G_1$. Hence $\textbf{f}$ is non-zero. Using the Schwartz-Zippel Lemma $\textbf{f}\neq 0$ and thus $\begin{bmatrix}T_1G_1& T_2\end{bmatrix}$ is invertible with probability at least $1-\NetSize/p$ over the choices of $\mathcal{X}$.
\end{IEEEproof}

\noindent \emph{Decoding:} The two message matrices $M_1$, $M_2$ along with the packets inserted by the adversary are transmitted to sink $\sink$ through the network with the use of random linear network coding (see Section~\ref{The_Random_Linear_Network_Coding}) and therefore sink $\sink$ gets:
\begin{align}
Y &= T_1 M_1 + T_2 M_2 + T_z Z\notag\\
\Leftrightarrow Y &= \begin{bmatrix}Y_1&Y_2&Y_3\end{bmatrix} =
\begin{bmatrix}T_1&T_2&A\end{bmatrix}+E,
\label{eqn:Y_Y1_Y2_Y3}
\end{align}
where $A = T_1M'_1 + T_2M'_2\in \F_p\hspace{-1.1mm}^{m_{\sources_1,\sources_2}\times knN}$
and $E\in\F_p\hspace{-1.1mm}^{m_{\sources_1,\sources_2}\times \ell}$ has rank no more than $z$ over field $\F_p$. Let $E=\begin{bmatrix} E_1&E_2&E_3\end{bmatrix}$, where $E_1\in \F_p\hspace{-1.1mm}^{m_{\sources_1,\sources_2}\times n}$, $E_2\in \F_p\hspace{-1.1mm}^{m_{\sources_1,\sources_2}\times N}$ and $E_3\in\F_p\hspace{-1.1mm}^{m_{\sources_1,\sources_2}\times  knN}$. Sink $\sink$ will first decode $M_2$ and then $M_1$.

\emph{Stage 1: Decoding $X_2$:}
Let $Y_a=\begin{bmatrix}Y_1G_1& Y_2 & Y_3^f\end{bmatrix}$ be a
matrix in $\F_q\hspace{-1mm}^{m_{\sources_1,\sources_2}\times(R_1+N+kN)}$. To be precise:
\begin{eqnarray}Y_a
& =&
 \begin{bmatrix}
  T_1G_1&T_2&A^f
 \end{bmatrix}
 +
 \begin{bmatrix}
  E_1G_1 & E_2 & E_3^f
 \end{bmatrix}.\label{eq:Y_a}
\end{eqnarray}
Sink $\sink$ uses invertible row operations over
$\F_{q}$ to transform $Y_a$ into a row-reduced echelon matrix
$\begin{bmatrix} T_{RRE}&M_{RRE}\end{bmatrix}$ that has the same row
space as $Y_a$, where $T_{RRE}$ has $m_{\sources_1,\sources_2}=R_1+N$ columns and $M_{RRE}$
has $kN$ columns. Then the following propositions are from the
results\footnote{1) is from Prop. 7, 2) from Thm. 9, and 3)
from Prop. 10 in~\cite{SKK08}.} proved
in~\cite{SKK08}:
\begin{proposition}
\begin{enumerate}
\item \label{subPro:RRE} The matrix $\begin{bmatrix} T_{RRE}&M_{RRE}\end{bmatrix}$ takes
the form
 $\begin{bmatrix} T_{RRE}&M_{RRE}\end{bmatrix}=\begin{bmatrix}I_{C}+\hat{L}{U}_\mu^T&r\\O&\hat{E}\end{bmatrix}$,
where ${U}_\mu\in \F_{q}\hspace{-1mm}^{C\times \mu}$ comprises of $\mu$ distinct
columns of the $C\times C$ identity matrix such that ${U}_\mu^Tr=0$
and ${U}_\mu^T\hat L=-I_{\mu}$. In particular, $\hat{L}$ in $\F_{q}\hspace{-1mm}^{C\times \mu}$ is the ``error-location matrix", $r\in\F_{q}\hspace{-1mm}^{C\times kN}$ is the ``message matrix", and $\hat E\in\F_{q}\hspace{-1.0mm}^{\delta \times kN}$ is the ``known error value" (and its rank is denoted $\delta$).
\item \label{subPro:Row-dis} Let  $X=\begin{bmatrix} X_1 \\
M_2^f\end{bmatrix}$ and $e=r-X$ and $\tau=\text{rank}\begin{bmatrix} \hat{L} & e\\
0 &\hat{E} \end{bmatrix}$. Then $2\tau-\mu-\delta$ is no more than
$d_S(\left\langle\begin{bmatrix} T_{RRE} & M_{RRE}
\end{bmatrix}\right\rangle,\left\langle\begin{bmatrix} I_{m_{\sources_1,\sources_2}}&X\end{bmatrix}\right\rangle)$, {\it i.e.}, the subspace distance between $\left\langle\begin{bmatrix}
T_{RRE} & M_{RRE}
\end{bmatrix}\right\rangle$ and $\left\langle\begin{bmatrix} I_{m_{\{\sources_1,\sources_2\}}}&X\end{bmatrix}\right\rangle$.

\item \label{subPro:errorform} There exist $\tau$ column vectors
$\mathbf L_1,\mathbf L_2,\ldots,\mathbf L_\tau\in\F_{q}\hspace{-1mm}^{C}$ and $\tau$ row vectors $\mathbf E_{1},\mathbf E_{2},\ldots,\mathbf E_{\tau}\in\F_{q}\hspace{-1mm}^{1\times kN}$ such that $e=\sum_{i\in [1,\tau]} \mathbf L_i \mathbf E_{i}$. In particular, $\mathbf L_1,\mathbf L_2,\ldots,\mathbf L_{\mu}$ are the columns of $\hat{L}$, and $\mathbf E_{\mu+1},\mathbf E_{\mu+2},\ldots,\mathbf E_{\mu+\delta}$
are the rows of $\hat{E}$.
\end{enumerate}
\label{pro:RRE}
\end{proposition}

In the following subscript $d$ stands for the last $N$ rows of any
matrix/vector. Then we show the following for our scheme.
\begin{lemma}
1) Matrix $\hspace{1mm} e_d=r_d-M_2^f$ can be expressed as
$e_d=\sum_{i\in 1,2,\ldots,\tau}(\mathbf L_i)_d \mathbf E_i$, where
$(\mathbf L_1)_d,(\mathbf L_2)_d,\ldots,(\mathbf L_{\mu})_d$ are the
columns of $\hat L_d$ and $\mathbf E_{\mu+1},\mathbf
E_{\mu+2},\ldots,\mathbf E_{\mu+\delta}$ are the rows of $\hat{E}$.

$2)\hspace{1mm}$With probability at least $1-\NetSize/p$, $2\tau-\mu-\delta\leq 2z$
\label{Le:CorrectX2}
\end{lemma}
\begin{IEEEproof}
$1)$ It is a direct corollary from the third statement of Proposition~\ref{pro:RRE}.

\noindent$2)$ Using the second statement of Proposition~\ref{pro:RRE} it suffices to prove with probability at least $1-\NetSize/p$, $d_S(\left\langle\begin{bmatrix} T_{RRE} & M_{RRE}\end{bmatrix}\right\rangle, \left\langle\begin{bmatrix} I_{m_{\sources_1, \sources_2}}&X\end{bmatrix}\right\rangle)\leq 2z$.

As shown in the proof of Lemma~\ref{lemma:RankBoundE}, the columns
of $E_3^f$ are in the column space of $E_3$ (and then of $E$) over
$\F_{q}$. Thus $\begin{bmatrix}E_1&E_2&E_3^f\end{bmatrix}$ and
therefore $\begin{bmatrix}E_1G_1&E_2&E_3^f\end{bmatrix}$ has rank at
most equal to $z$ over $\F_q$. Using
Proposition~\ref{pro:rowdis-rankdis} and (\ref{eq:Y_a}),
$d_S(\left\langle Y_a\right\rangle,\left\langle\begin{bmatrix}T_1G_1&T_2&A^f\end {bmatrix}\right\rangle)$ is no more
than $2z$. Since $d_S(\left\langle\begin{bmatrix} T_{RRE} & M_{RRE}\end{bmatrix}\right\rangle,\left\langle Y_a\right\rangle)\\=0$,
we have $d_S(\left\langle\begin{bmatrix} T_{RRE} &
M_{RRE}\end{bmatrix}\right\rangle,\left\langle\begin{bmatrix}T_1G_1&T_2&A^f\end
{bmatrix}\right\rangle)\leq 2z$.

Using Lemma~\ref{lemma:Invertible_D}, matrix $D=\begin{bmatrix}T_1G_1& T_2\end{bmatrix}$ is invertible with
probability at least $1-\NetSize/p$, so $\begin{bmatrix} I_{m_{\sources_1,\sources_2}}&X\end
{bmatrix}$ has zero subspace distance from $\begin{bmatrix}
D&DX\end {bmatrix}=\begin{bmatrix}T_1G_1&T_2&A^f\end {bmatrix}$.
Thus,
\begin{align*}
d_S(\left\langle\begin{bmatrix} T_{RRE} &
M_{RRE}\end{bmatrix}\right\rangle,\left\langle\begin{bmatrix} I_{m_{\sources_1,\sources_2}}&X\end{bmatrix}\right\rangle)\leq 2z.
\end{align*}
\end{IEEEproof}

In the end combining Lemma~\ref{Le:CorrectX2} and
Theorem~\ref{Th:GabiDecode} sink $\sink$ can take
$(\hat L_d, \hat E, r)$ as the input for the Gabidulin decoding
algorithm and decode $X_2$ correctly.

\emph{Stage 2: Decoding ${X_1}$:}
From (\ref{eqn:Y_Y1_Y2_Y3}) sink $\sink$ gets
$Y=\begin{bmatrix}T_1+E_1&T_2+E_2&A+E_3\end{bmatrix}$,
computes $(T_2+E_2)M_2$, and then subtracts matrix
$\begin{bmatrix}O&(T_2+E_2)& (T_2+E_2)M_2\end{bmatrix}$ from $Y$.
The resulting matrix has $N$ zero columns in the middle (column
$n+1$ to column $n+N$). Disregarding these we get:
\begin{eqnarray*}
Y' = \begin{bmatrix}T_1&T_1 M_1\end{bmatrix}+
\begin{bmatrix}E_1&E_3-E_2 M_2\end{bmatrix}.
\end{eqnarray*}

The new error matrix $E' =
\begin{bmatrix}E_1&E_3-E_2 M_2\end{bmatrix}$ has rank at
most $z$ over $\F_p$ since the columns of $E'$ are simply linear
combinations of columns of $E$ whose rank is at most $z$. Therefore
the problem degenerates into a single source problem and sink
$\sink$ can decode $X_1$ with probability at least
$1-\NetSize/p$ by following the approach in
\cite{SKK08}.

Summarizing the above decoding scheme for $X_1$ and $X_2$, we have the following main result:
\begin{theorem}
Each $\sink$ can efficiently decode the information from all sources correctly with probability at least $1-|\SourceNum|\NetSize/p$.
\label{Th:non-coherent}
\end{theorem}

\emph{Decoding complexity:} For both coherent and non-coherent cases the computational complexity of Gabidulin encoding and decoding of two source messages is dominated by the decoding of $X_2$, which requires $\mathcal O(nNm_\sources\ell  \log(p n N))$ operations over $\F_p$
(see~\cite{SKK08}).

To generalize our technique to more sources, consider a network with
$\SourceNum$ sources $\mathcal {S}_1,\mathcal {S}_2,\ldots,\mathcal
{S}_\SourceNum$. Let $R_i$ be the rate of $\mathcal {S}_i$ and
$n_i=R_i+2z$ for each $i\in[1,\SourceNum]$. A straightforward
generalization uses the multiple-field-extension technique so that
$\mathcal{S}_i$ uses the generator matrix over finite field of size
$p^{n_1 n_2\ldots n_i}$. In the end the packet length must be at
least $n_g=n_1 n_2\ldots n_\SourceNum$, resulting in a decoding
complexity $\mathcal O(m_\sources n_g^2 \log(p n_g))$ increasing exponentially
in the number of sources $\SourceNum$. Thus the multiple
field-extension technique works in polynomial time only for a fixed
number of sources.

Note that the intermediate nodes work in the base field $\F_p$ to perform random linear network coding. The multiple-field-extension is an end-to-end technique, {\it i.e.}, only the sources and sinks use the extended field.

\subsection{Coherent case}
Sections~\ref{The_Random_Secret_Model},~\ref{General_approach} and
~\ref{Polynomial_time_construction} give code constructions for the
non-coherent coding scenario. Note that  a non-coherent coding scheme can also be applied in the coherent setting when the network is known.  Hence, the capacity regions of coherent and non-coherent network coding for the same multi-source multicast network are the same. However, both the constructions of Sections ~\ref{General_approach} and ~\ref{Polynomial_time_construction} include an overhead of
incorporating a global coding vector. Therefore, they achieve the
outer bounds given by (\ref{eq_non-coherent}) only asymptotically in packet length. In contrast, in the coherent case, the full capacity region can be achieved exactly with packets of finite length, as shown in the following:
\begin{IEEEproof}[Proof of Theorem~\ref{n_sources}, coherent case achievability]
We first construct a multi-source multicast network code $\cC$ for $\graph$ that can correct any $2z$ errors with known locations,  called erasures in~\cite{yang07characterizations}. We can use the result of~\cite{dana06capacity} for multi-source multicast network coding in an alternative model where on each link either an erasure symbol or error-free information is received, by observing the following correspondence between the two models. We form a graph $\graph'$ by replacing each link $l$ in $\graph$ with two links in tandem  with a new node $v_l$ between them, and adding an additional source node $ u$ of rate $2z$ connected by a new link $k_l$ to each node $v_l$. We use the result from~\cite{dana06capacity} to obtain a multi-source network code that achieves a given rate vector under any pattern of erasure symbols such that  the maxflow-mincut conditions are satisfied for every subset of sources in $\graph'$. In particular, if erasure symbols (by the definition of~\cite{dana06capacity}) are received  on all but $2z$ of the new links $k_l$ (corresponding to $2z$ erasures in $\graph $ by the definition of~\cite{yang07characterizations}), all the original sources can be decoded.

Let $l_{i,j},j = 1,\dots,n_i,$ be the outgoing links of each source $s_i,i = 1, \ldots, n$.  Next, we construct the graph $\graph_{\sources}$ from $\graph$ by adding  a virtual super source node $w$, and $n_i$ links $l'_{i,j},j = 1,\dots,n_i,$ from $w$ to each source $s_i$. Then the code $\cC$ for the multi-source problem corresponds to a single-source network code $\cC_{\sources}$ on $\graph_{\sources}$ where the symbol on each link $l'_{i,j}$ is the same as that on link $l_{i,j}$, and the coding operations at all other nodes are identical for $\graph_{\sources'}$ and $\graph_{\sources}$.

By~\cite{yang07characterizations} the following are equivalent in the single-source case:
\begin{enumerate}
\item a linear network code has network minimum distance at least $2z +1$
\item the code corrects any error of weight at most $z$
\item the code corrects any erasure of weight at most $2z$.
\end{enumerate}
This implies that $\cC_{\sources}$ has network minimum distance at least $2z+1$, and so it can correct any $z$ errors.
\end{IEEEproof}

\section{Extension to more than two sources}
\label{more_than_two_sources}

When there are more than two sources the extension of our encoding and decoding techniques is straightforward both for the case of the side-channel and the omniscient model, and up to this point we have focused on the case of two sources simply for notational convenience. To clarify how our techniques can extend to multiple sources we will outline the encoding and decoding for an arbitrary number of sources equal to $\SourceNum$ and use results from the previous sections.

\noindent\emph{Side-channel model}: For the case of the side-channel model each source encodes its data $X_i\in\F_p\hspace{-1.1mm}^{R_i\times(\ell-\alpha)}$, $i\in\{1,\ldots,s\}$,  in a matrix $M_i=\begin{bmatrix}L_i &X_i\end{bmatrix}$ where $L_i\in\F_p\hspace{-1.1mm}^{R_i\times\alpha}$ will be such so that equation $H_i=M_iP_i$ holds. Source $\sources_i$ shares with the receiver/receivers the random matrix $H_i\in\F_p\hspace{-1.1mm}^{R_i\times\alpha}$ along with the random vector $W_i = \begin{bmatrix}r_{i1}&r_{i2}&\ldots&r_{i\alpha}\end{bmatrix}$. The vector $W_i$ defines matrix $P_i\in\F_p\hspace{-1.1mm}^{\ell\times\alpha}$ since its $(m,n)-\text{th}$ entry equals $(r_{in})^m$. Every receiver follows the decoding steps described in Section~\ref{The_Random_Secret_Model} and gets equations $M_iP_i=M_i^s(FH_i)=H_i$, $i\in\{1,\ldots,s\}$, that can be solved with high probability using Gaussian elimination.

\noindent\emph{Omniscient model}: For the case of the omniscient adversary we will need to extend the field we work with $\SourceNum$ times. Assume that $n_i=R_i+2z$ and the information from source $\sources_i$ is organized into a matrix $X_i\in\F_p\hspace{-1.1mm}^{R_i\times kn_1\ldots n_\SourceNum}$. Before transmission matrix $X_i$, $i\in\{1,\ldots,\SourceNum-1\}$, is viewed as matrix $X_i\in\F_{p_i}\hspace{-1.1mm}^{R_i\times kn_{i+1}\ldots n_\SourceNum}$ in the larger field $\F_{p_i}$ where $p_i=p^{n_1\ldots n_i}$ and $X_\SourceNum$ is viewed as a matrix $X_\SourceNum\in\F_{p_\SourceNum}\hspace{-1.1mm}^{R_\SourceNum\times k}$ where $p_\SourceNum=p^{n_1\ldots n_\SourceNum}$. Each matrix $X_i$ is multiplied with a generator matrix $G_i\in\F_{p_i}\hspace{-1.1mm}^{n_i\times R_i}$, creating $G_iX_i$ whose unfolded version $M_i'=(G_iX_i)^u$ is a matrix in $\F_p\hspace{-1.1mm}^{n_i\times kn_1\ldots n_\SourceNum}$. All matrices $G_i$ are chosen as generator matrices for Gabidulin codes and have the capability of correcting errors of rank at most $z$ over field $\F_{p_i}$.

Source $\sources_1$ create the message matrix $M_1$ by appending some header to $M_1'$, specifically the message is $M_1=\begin{bmatrix}I_{n_1}&O_{n_1\times n_2}&\ldots&O_{n_1\times n_\SourceNum}&M_1'\end{bmatrix}$ where $I_{n_1}$ is the identity matrix with dimensions $n_1\times n_1$ and $O_{n_i\times n_j}$ is the zero matrix with dimensions $n_i\times n_j$. Similarly $M_2=\begin{bmatrix}O_{n_2\times n_1}\ I_{n_2}&\ldots&O_{n_2\times n_\SourceNum}&M_2'\end{bmatrix},\ldots,\\ M_\SourceNum=\begin{bmatrix}O_{n_\SourceNum\times n_1}\ O_{n_\SourceNum\times n_2}&\ldots&I_{n_\SourceNum}&M_\SourceNum'\end{bmatrix}$ and therefore the packet length is $\ell=\sum_{i=1}^\SourceNum n_i+k\prod_{i=1}^\SourceNum n_i$ over $\F_p$ the base field of network coding.

Similar to equation (\ref{eqn:Y_Y1_Y2_Y3}) the received matrix can be written as
\begin{align*}
&Y = T_1M_1+\ldots+T_\SourceNum M_\SourceNum + T_z Z\\
\Leftrightarrow&Y = \begin{bmatrix}Y_1&\ldots&Y_\SourceNum&Y_{\SourceNum+1}\end{bmatrix} =\begin{bmatrix}T_1&\ldots&T_\SourceNum&A'\end{bmatrix}+E
\end{align*}
where $A'=T_1M_1'+\ldots+T_\SourceNum M_\SourceNum '$ and $E\in\F_p\hspace{-1.1mm}^{m_\sources\times \ell}$ has rank no more than $z$ over field $\F_p$. For the decoding of information from source $\sources_\SourceNum$ we form the matrix $Y_\alpha '=\begin{bmatrix}Y_1G_1&\ldots& Y_{\SourceNum-1}G_{\SourceNum-1}& Y_\SourceNum&Y_{\SourceNum+1}^f\end{bmatrix}$ and transform it to a row-reduced echelon form as in Proposition~\ref{pro:RRE}. Since matrix $D'=\begin{bmatrix}T_1G_1&\ldots& T_{\SourceNum-1}G_{\SourceNum-1}&T_\SourceNum\end{bmatrix}$ is invertible with high probability similar to Lemma~\ref{lemma:Invertible_D} one can use Lemma~\ref{Le:CorrectX2} and decode $X_\SourceNum$. By subtracting $\begin{bmatrix}O_{m_\sources\times n_1}&\ldots&O_{m_\sources\times n_{\SourceNum-1}}&Y_\SourceNum&Y_\SourceNum M_\SourceNum '\end{bmatrix}$ from $Y$ the problem reduces to $\SourceNum-1$ number of sources and one can solve it recursively.

\section{Comparison of our code constructions}
{\small{
\begin{table}[ht]
\caption{Comparison of performance metrics of the code constructions given in Sections~\ref{The_Random_Secret_Model},~\ref{General_approach} and ~\ref{Polynomial_time_construction} for any achievable rate vector $(R_1, R_2, \ldots, R_s)$}
\centering 
\begin{tabular}{|c | c | c |}
\hline
   & decoding complexity & packet length \\
\hline 
Side-channel model &$O(\ell m_\sources^3)$& $\Theta(m_\sources^2)$\\
\hline 
Omniscient adversary: &$O(p^{lm_\sources})$& $\Theta(m_\sources)$\\
subspace codes &&\\
\hline
Omniscient adversary: &$O(m_\sources^{2s+1}\log(pm_\sources^s))$&$\Theta(\prod_{i=1}^s  m_{\sources_i})$\\
field extension codes&&\\
\hline
\end{tabular}
\label{table_comparison}
\end{table}}}
In this section we compare some performance metrics of the code constructions given in Sections~\ref{The_Random_Secret_Model},~\ref{General_approach} and ~\ref{Polynomial_time_construction}. For convenience, Table~\ref{table_comparison} summarizes the requirements on the decoding complexity and the packet length for each of the achievable schemes. For clarity of comparison, we approximate all quantities presented in Table~\ref{table_comparison}; the exact expressions are derived in
the corresponding sections.

Based on Table~\ref{table_comparison}, we can make the following observations about the practicality of our constructions:
\begin{itemize}
\item If the secret channel is available, one should use the side-channel model construction since it not only achieves higher rates but also provides lower decoding complexity.
\item Multiple-field extension codes have computational complexity that is polynomial in all network parameters, but exponential in the number of sources. Therefore, they are preferable when the number of sources is small.
\item Random subspace codes become beneficial compared to multiple-field extension codes as the number of sources grows.
\end{itemize}

\section{Conclusion}
In this work we consider the problem of communicating messages from multiple sources to multiple sinks over a network that contains a hidden malicious adversary who observes and attempts to jam communication. We consider two models. In the first model, the sources share a small secret (that is unknown to the adversary) with the sink(s). In the second model, this resource is unavailable -- no limitations on the adversary's knowledge are assumed. We prove upper bounds on the set of achievable rates in these settings. Since more resources are available to the honest parties in the first model, the rate-region corresponding to the upper bounds in the first model is larger than that in the second model. We also provide novel algorithms that achieve any point in the rate-regions corresponding to the two models. Our codes for the first model have computational complexity that is polynomial in network parameters. For the second model we have two algorithms.
In our codes based on random subspace design, all sources code over the same field, and decoding is based on minimum injection distance.
Our codes based on multiple-field extension have computational complexity that is polynomial in all network parameters, but exponential in the number of sources.

Our codes are end-to-end and decentralized -- each interior node is oblivious to the presence of an adversary, and merely performs random linear network coding. They also do not require prior knowledge of the network topology or coding operations by any honest party. They work in the presence of a computationally unbounded adversary, even one who knows the network topology and coding operations and can decide where and how to jam the network on the basis of this information.

A problem that remains open is that of computationally efficient codes for the omniscient adversarial case with a large number of sources. This may require new insights in algebraic code design.

Besides multi-source multicast, our codes have implications for the much more common scenario of multiple unicasts. One class of codes (that is not rate-optimal) for this problem assumes that each sink treats information that it is uninterested in as noise, and decodes and successively cancels such messages out. Since the code constructions provided here achieve higher rates than those available in prior work, they may aid in non-trivial achievability schemes (though in general still not rate-optimal) for this problem.

\bibliography{lit}

\begin{thebibliography}{10}
\providecommand{\url}[1]{#1}
\csname url@samestyle\endcsname
\providecommand{\newblock}{\relax}
\providecommand{\bibinfo}[2]{#2}
\providecommand{\BIBentrySTDinterwordspacing}{\spaceskip=0pt\relax}
\providecommand{\BIBentryALTinterwordstretchfactor}{4}
\providecommand{\BIBentryALTinterwordspacing}{\spaceskip=\fontdimen2\font plus
\BIBentryALTinterwordstretchfactor\fontdimen3\font minus
  \fontdimen4\font\relax}
\providecommand{\BIBforeignlanguage}[2]{{%
\expandafter\ifx\csname l@#1\endcsname\relax
\typeout{** WARNING: IEEEtran.bst: No hyphenation pattern has been}%
\typeout{** loaded for the language `#1'. Using the pattern for}%
\typeout{** the default language instead.}%
\else
\language=\csname l@#1\endcsname
\fi
#2}}
\providecommand{\BIBdecl}{\relax}
\BIBdecl

\bibitem{ACLY00}
R.~Ahlswede, N.~Cai, S.-Y.~R. Li, and R.~W. Yeung, ``Network information
  flow,'' \emph{IEEE Trans.~Inf.~Theory}, vol.~46, no.~6, pp. 1204--1216, Jul.
  2000.

\bibitem{LYC03}
S.-Y.~R. Li, R.~W. Yeung, and N.~Cai, ``Linear network coding,'' \emph{IEEE
  Trans.~Inf.~Theory}, vol.~49, no.~2, pp. 371--381, Feb. 2003.

\bibitem{KM03}
R.~K\"otter and M.~M{\'e}dard, ``An algebraic approach to network coding,''
  \emph{IEEE/ACM Trans. on Networking}, vol.~11, no.~5, pp. 782--795, Oct.
  2003.

\bibitem{Ho_etal06}
T.~Ho, M.~M{\'e}dard, R.~K\"otter, D.~R. Karger, M.~Effros, J.~Shi, and
  B.~Leong, ``A random linear network coding approach to multicast,''
  \emph{IEEE Trans.~Inf.~Theory}, vol.~10, no.~52, pp. 4413--4430, Oct. 2006.

\bibitem{RaymondNECC2002}
N.~Cai and R.~W. Yeung, ``Network coding and error correction,'' in
  \emph{Proc.\ of 2002 IEEE Information Theory Workshop (ITW)}, 2002.

\bibitem{YC06}
------, ``Network error correction, part {I}: {B}asic concepts and upper
  bounds,'' \emph{Commun.~Inf.~Syst}, vol.~6, no.~1, pp. 19--36, 2006.

\bibitem{CY06}
R.~W. Yeung and N.~Cai, ``Network error correction, part {II}: {L}ower
  bounds,'' \emph{Commun.~Inf.~Syst}, vol.~6, no.~1, pp. 37--54, 2006.

\bibitem{SKK08}
D.~Silva, F.~R. Kschischang, and R.~K\"otter, ``A rank-metric approach to error
  control in random network coding,'' \emph{IEEE Trans.~Inf.~Theory}, vol.~54,
  no.~9, pp. 3951--3967, Sep. 2008.

\bibitem{Jaggi_etal08}
S.~Jaggi, M.~Langberg, S.~Katti, T.~Ho, D.~Katabi, M.~M{\'e}dard, and
  M.~Effros, ``Resilient network coding in the presence of {B}yzantine
  adversaries,'' \emph{IEEE Trans.~Inf.~Theory}, vol.~54, no.~6, pp.
  2596--2603, Jun. 2008.

\bibitem{Roth06}
R.~M. Roth, \emph{Introduction to Coding Theory}.\hskip 1em plus 0.5em minus
  0.4em\relax Cambridge University Press, 2006.

\bibitem{KK08}
R.~K\"otter and F.~R. Kschischang, ``Coding for errors and erasures in random
  network coding,'' \emph{IEEE Trans.~Inf.~Theory}, vol.~54, no.~8, pp.
  3579--3591, Aug. 2008.

\bibitem{SFD08}
M.~J. Siavoshani, C.~Fragouli, and S.~Diggavi, ``Noncoherent multisource
  network coding,'' in \emph{Proc.~IEEE Int.~Symposium on Inform.~Theory},
  Toronto, Canada, Jul. 2008, pp. 817--821.

\bibitem{FragouliITW09MultiSource}
S.~Mohajer, M.~Jafari, S.~Diggavi, and C.~Fragouli, ``On the capacity of
  multisource non-coherent network coding,'' in \emph{Proc.\ of the IEEE
  Information Theory Workshop}, 2009.

\bibitem{FragouliMibiHoc09MultiSource}
M.~Siavoshani, C.~Fragouli, and S.~Diggavi, ``Code construction for multiple
  sources network coding,'' in \emph{Proc.\ of the MobiHoc}, 2009.

\bibitem{fragouli_personal}
Personal Communication.

\bibitem{landberg08}
L.~Nutman and M.~Langberg, ``Adversarial models and resilient schemes for
  network coding,'' in \emph{Proc.\ of IEEE International Symposium of
  Information Theory}, 2008, pp. 171--175.

\bibitem{RandCode0}
T.~Ho, M.~M{\'e}dard, J.~Shi, M.~Effros, and D.~R. Karger, ``On randomized
  network coding,'' in \emph{Proc. of Allerton 2003}, 2003.

\bibitem{Svit_rate_regions}
S.~Vyetrenko, T.~Ho, M.~Effros, J.~Kliewer, and E.~Erez, ``Rate regions for
  coherent and noncoherent multisource network error correction,'' in
  \emph{Proc.\ of IEEE International Symposium of Information Theory}, 2009.

\bibitem{hongyi_ted}
H.~Yao, T.~K. Dikaliotis, S.~Jaggi, and T.~Ho, ``Multiple access network
  information-flow and correction codes$^*$,'' in \emph{Proc.\ of IEEE
  Information Theory Workshop, Dublin}, 2010.

\bibitem{Algebra_Martin}
M.~Artin, \emph{Algebra}.\hskip 1em plus 0.5em minus 0.4em\relax New Jersey:
  Prentice Hall, 1991.

\bibitem{sv09noncoherent}
S.~Vyetrenko, T.~Ho, and E.~Erez, ``On noncoherent correction of network errors
  and erasures with random locations,'' in \emph{Proc.\ of the IEEE
  International Symposium on Information Theory}, Jun. 2009.

\bibitem{silvametric}
D.Silva and F.~R. Kschischang, ``On metrics for error correction in network
  coding,'' \emph{IEEE Transactions on Information Theory}, vol.~55, pp.
  5479--5490, 2009.

\bibitem{Gabidulin_TheoryOfCodes_1985}
E.~M. Gabidulin, ``Theory of codes with maximum rank distance,'' \emph{Probl.
  Peredachi Inf.}, vol.~21, no.~1, pp. 3--16, 1985.

\bibitem{CCBook}
M.~Agrawa and S.~Biswas, ``Primality and identity testing via chinese
  remaindering,'' \emph{Journal of the ACM}, 2003.

\bibitem{yang07characterizations}
S.~Yang and R.~W. Yeung, ``Characterizations of network error
  correction/detection and erasure correction,'' in \emph{NetCod 2007}, Jan
  2007.

\bibitem{dana06capacity}
A.~F. Dana, R.~Gowaikar, R.~Palanki, B.~Hassibi, and M.~Effros, ``Capacity of
  wireless erasure networks,'' \emph{IEEE Transactions on Information Theory},
  vol.~52, pp. 789--804, 2006.

\end{thebibliography}
\bibliographystyle{IEEEtran}
\end{document}